\documentclass[notitlepage,a4paper,aps,prd,superscriptaddress,nofootinbib,groupedaddress,twocolumn]{revtex4}
\usepackage{color}
\usepackage{textcomp}
\usepackage{amsmath}
\usepackage{amssymb}
\usepackage{amsfonts}
\usepackage[utf8]{inputenc}
\usepackage{subcaption}
\usepackage{graphicx}
\usepackage{hyperref}
\usepackage{multirow}
\usepackage{accents}
\usepackage[english]{babel}

\makeatletter

\@ifundefined{textcolor}{}
{%
 \definecolor{BLACK}{gray}{0}
 \definecolor{WHITE}{gray}{1}
 \definecolor{RED}{rgb}{1,0,0}
 \definecolor{GREEN}{rgb}{0,1,0}
 \definecolor{BLUE}{rgb}{0,0,1}
 \definecolor{CYAN}{cmyk}{1,0,0,0}
 \definecolor{MAGENTA}{cmyk}{0,1,0,0}
 \definecolor{YELLOW}{cmyk}{0,0,1,0}
}

\usepackage{bm}
\usepackage{indentfirst}
\usepackage{float}
\usepackage{caption}
\usepackage{psfrag}
\usepackage{epstopdf}
\usepackage{diagbox}
\usepackage{orcidlink}

\makeatother
\begin{document}

\title{Constraining Lorentz Invariance Violation via Precision Atomic Spectroscopy}
\author{A. S. Lemos}
\email{adiel@ufersa.edu.br}
\affiliation{Departamento de Ci\^{e}ncias Exatas e Tecnologia da Informa\c{c}\~{a}o, Universidade
Federal Rural do Semi-\'{A}rido, 59515-000 Angicos, Rio Grande do Norte, Brazil}
\author{J. E. J. Matias}
\email{joaoeliasfisica1@yahoo.com}
\affiliation{Physics Department, Federal University of Para\'iba, Caixa Postal 5008, 58059-900, Jo\~ao Pessoa-PB, Brazil}
\author{F. Dahia}
\email{fdahia@fisica.ufpb.br}
\affiliation{Physics Department, Federal University of Para\'iba, Caixa Postal 5008, 58059-900, Jo\~ao Pessoa-PB, Brazil}
\author{I. P. Lobo}
\email{iarley.lobo@academico.ufpb.br}
\affiliation{Physics Department, Federal University of Para\'iba, Caixa Postal 5008, 58059-900, Jo\~ao Pessoa-PB, Brazil}
\affiliation{Physics Department, Federal University of Campina Grande, Caixa Postal 10071, 58429-900, Campina Grande-PB, Brazil}

\begin{abstract}
This paper investigates Lorentz violation arising from quantum gravity effects in electronic and muonic hydrogen atoms. The violation is modeled through a modified dispersion relation parameterized by a dimensionless coefficient \(\xi\) and suppressed by the Planck scale. Using high-precision spectroscopic data and a chi-squared fitting procedure that carefully accounts for theoretical uncertainties in the Rydberg constant and the proton radius, the analysis yields a 95\% confidence level constraint \(\xi \in [-14.3,\; 1.37] \times 10^5\). The mass scaling of the effect reveals two distinct regimes: for the lowest-order deformation, the heavy nucleus amplifies the signal, while for higher-order corrections, the lepton dominates the sensitivity. These bounds approach Planck-scale sensitivity using the simplest atomic systems, thereby circumventing ambiguities inherent in more complex quantum systems.
\end{abstract}

\maketitle

\section{Introduction}
A fundamental challenge in modern physics is the search for observable signatures of quantum gravity (QG), whose effects are suppressed by the immense Planck scale \(M_{\text{Pl}} \approx 1.22 \times 10^{19}\, \text{GeV}\). Despite this, innovative phenomenological approaches, enhanced by amplifying mechanisms such as cosmological propagation or those found when applying conservation laws, and by the extraordinary precision of modern experiments, suggest that such signatures may still be within reach \cite{Amelino-Camelia:2008aez}.

One prominent line of inquiry explores the possible violation or deformation of Lorentz symmetry \cite{Jacobson:2005bg,Amelino-Camelia:2000stu,Magueijo:2001cr}. While traditionally investigated in high-energy astrophysical settings \cite{Addazi:2021xuf,AlvesBatista:2023wqm}, recent attention has turned increasingly toward ultraviolet/infrared (UV/IR) mixing \cite{Amelino-Camelia:2008aez,Berglund:2022qcc}. This mechanism enables Planck-scale corrections to manifest even in low-energy regimes, leaving signatures on the quantum systems, potentially influencing spacetime symmetries \cite{Ballesteros:2025ypr,Ballesteros:2021dob,Arzano:2014cya,Arzano:2022nlo,Lobo:2025vqt}, dispersion relations \cite{Amelino-Camelia:2009wvc,Wagner:2023fmb,Varao:2024eig,Amelino-Camelia:2025zkq}, and uncertainty principles \cite{Bosso:2023aht}. Such low-energy effects become accessible in contexts where high precision or intrinsic amplification enhances sensitivity to Planck-scale physics—examples include cold-atom systems \cite{Amelino-Camelia:2009wvc}, studies of decoherence \cite{Arzano:2014cya,Petruzziello:2020wkd,Arzano:2022nlo,Lobo:2025mze,Donadi:2024amp,Lobo:2025vqt}, and atomic transitions \cite{Balachandran:2010xk,Moumni:2013yta}.

The more complex the system, the harder it is to isolate quantum gravitational effects from others. For this reason, we considered the simplest atomic system, which is the hydrogen atom. Although similar systems have been studied previously \cite{Pu:2019lmo}, as the theoretical and experimental uncertainties are not treated rigorously, the very strict constraint found for the deformed Hamiltonian is not realistic. Additionally, the generalized uncertainty principle has been applied to this system \cite{Bosso:2023aht}, but it constrains only second-order effects in the quantum gravity scale, which are too faint to be detectable at the Planck scale. A further novelty of our approach is the use of a muonic atom. As we demonstrate, the Hamiltonian perturbation is sensitive to the system's mass, thereby opening new phenomenological opportunities in this domain. Our analysis also carefully considers the theoretical and experimental uncertainties that surround atomic spectroscopy in order to set a robust constraint on Lorentz violation using tabletop experiments.

Our paper is organized as follows. In Section~\ref{sec:theory}, we establish the fundamental theoretical framework by constructing the Lorentz-violating two-body Hamiltonian and deriving its general perturbative energy shifts. In Section~\ref{sec:dataset}, we present the spectroscopic datasets and detail the energy shifts induced by the modified dispersion relations in the specific levels of electronic and muonic hydrogen, establishing the comprehensive theoretical expressions required for the parameter adjustments. In Section~\ref{sec:bounds}, we present the constraints from this global adjustment and discuss the resulting physical picture. Finally, in Section~\ref{sec:conc}, we offer our concluding remarks.

\section{The Perturbed Hamiltonian}\label{sec:theory}

The kinematics of particles that propagate in a quantum spacetime can be effectively described by modifications of the dispersion relation $E^2-c^2p^2=M^2c^4+2\xi_{n}M_{\rm Pl}^{-1}c^{n+1}\sum_{n}M^{n}p^{3-n}$, whose correction is suppressed by an energy scale $M_{\rm Pl}$ that governs how the test particle probes the quantum spacetime, which is supposedly the Planck energy $\sim 1.22 \times 10^{19}$ GeV \cite{Amelino-Camelia:2008aez,Amelino-Camelia:2009wvc,Wagner:2023fmb,Lobo:2025mze}. For instance, from a Lagrangian formulation, this can be described by an effective theory approach beyond the Standard Model that models related to Lorentz Invariance Violation (LIV) \cite{Colladay:1998fq}.


Corrections of the relativistic dispersion relation can leave imprints even in the non-relativistic limit, in which the speed of light scale is much larger than the typical velocity scale of particles. At the quantum level, the non-relativistic Hamiltonian reflects in a modified Schrödinger equation, as derived and explored in \cite{Wagner:2023fmb,Varao:2024eig,Lobo:2025mze}, such that the leading-order correction to the Hamiltonian for a particle of mass $M$ can be parameterized as:
\begin{equation}
    \hat{H}_{\text{LIV}}^{(n)} = \xi_n \frac{(Mc)^{2-n}}{M_{\rm Pl}} |\hat{\mathbf{p}}|^n,
    \label{eq:Hamiltonian_Final}
\end{equation}
where $\hat{\mathbf{p}}$ is the momentum operator and $\xi_n$ is a dimensionless parameter of order unity. Here, we identify the suppression scale directly with the fundamental mass scale of the model.

To rigorously describe the effects of LIV in an atomic system, we must treat the atom as a two-body composite system consisting of a nucleus (mass $M_{\rm nuc}$) and a lepton (mass $m_\ell$). The total LIV modification to the Hamiltonian is the sum of the individual contributions from both constituents. In the laboratory frame, this reads:
\begin{equation}\label{eq:nucfermion} \hat{H}_{\text{LIV}}^{(n)} = \frac{\xi_n}{M_{\rm Pl}} \left[ (M_{\text{nuc}} c)^{2-n} |\hat{\mathbf{p}}_{\text{nuc}}|^n + (m_{\ell} c)^{2-n} |\hat{\mathbf{p}}_{\ell}|^n \right]. \end{equation}
Moving to the center of mass (CM) frame, where $\hat{\mathbf{p}}_{nuc} = -\hat{\mathbf{p}}_{\ell} \equiv \hat{\mathbf{p}}$, the operator simplifies:
\begin{equation}
\hat{H}_{\text{LIV}}^{(n)} = \frac{\xi_n \, c^{2-n}}{M_{\rm Pl}} \sum_{i=\text{nuc},\ell} \left( m_i^{2-n} \right) |\hat{\mathbf{p}}|^n.
\end{equation}

The mass scaling of the MDR leads to three distinct phenomenological regimes depending on the deformation order $n$. In the infrared regime ($n=1$), the Hamiltonian scales linearly with the particle mass ($\hat{H} \propto m_{i} |\mathbf{p}|$). Consequently, the contributions in the CM frame sum linearly, resulting in an amplification effect driven by the heavy nucleus:
\begin{equation}\label{eq:pertnuc}
    \hat{H}_{\text{LIV}}^{(1)} = \frac{\xi_1 c}{M_{\rm Pl}} (M_{\text{nuc}} + m_\ell) |\hat{\mathbf{p}}| \approx  \frac{\xi_1 c}{M_{\rm Pl}} M_{\text{nuc}} |\hat{\mathbf{p}}|.
\end{equation}
In this scenario, the atom acts as a macroscopic probe, where the sensitivity is enhanced by a factor of $M_{\text{proton}}/m_e \approx 1836$ compared to single-particle electron models. Here the proton is treated as an effective point-like particle, which is appropriate at atomic momentum scales, where its internal structure is not resolved; the relation between effective and constituent-level coefficients is part of the soccer-ball problem discussed in Sec.~\ref{subsec:macroscopic_scaling}.

In Eq.~\eqref{eq:nucfermion} we have assigned the same parameter $\xi_n$ to the nucleus and to the lepton. In a general framework these constituents may be characterized by different coefficients. However, if the modification of the dispersion relation originates from the quantum structure of spacetime, the coefficients of different particles are expected to differ at most by factors of order unity. For composite bodies, an important source of such differences is the number of constituents, as discussed in Sec.~\ref{subsec:macroscopic_scaling} in connection with the soccer-ball problem. In the hydrogen atom the nucleus is a single proton, so that nucleus and lepton are both individual particles, and no enhancement or suppression associated with a large number of constituents is expected to differentiate their coefficients. This is one of the main advantages of using the simplest atomic system, in contrast with heavier atoms whose nuclei contain many nucleons. Allowing for independent coefficients $\xi_1^{(p)}$ and $\xi_1^{(\ell)}$, Eq.~\eqref{eq:pertnuc} reads
\begin{equation}
\hat H^{(1)}_{\rm LIV}=\frac{c}{M_{\rm Pl}}\left(\xi_1^{(p)}M_{\rm proton}+\xi_1^{(\ell)}m_\ell\right)|\hat{\bf p}|,
\end{equation}
and the parameter constrained in Sec.~IV must be understood as the effective combination
\begin{equation}
\xi=\xi_1^{(p)}+\frac{m_\ell}{M_{\rm proton}}\,\xi_1^{(\ell)} .
\label{eq:xi_eff}
\end{equation}
If both coefficients are of the same order of magnitude, the leptonic term in electronic hydrogen is suppressed by $m_e/M_{\rm proton}\simeq5\times10^{-4}$, and our bound applies to $\xi_1^{(p)}$ up to corrections of this size. In muonic hydrogen the same combination holds with $m_\mu/M_{\rm proton}\simeq0.113$, so the leptonic term is far less suppressed. The muonic data nevertheless constrain only $r_p$ in our adjustment, since the MDR contribution to the muonic Lamb shift is of order $1$~kHz for $\xi\sim10^{6}$, well below its experimental uncertainty of $5.6\times10^{5}$~kHz.

A qualitative shift occurs in the quadratic case ($n=2$), which represents a mass-invariant regime. Here, the mass dependence vanishes entirely since $(mc)^{2-2} = 1$, and the correction mimics a kinetic energy renormalization that affects both particles equally, regardless of their mass hierarchy:
\begin{equation}
    \hat{H}_{\text{LIV}}^{(2)} = \frac{\xi_2}{M_{\rm Pl}} (1 + 1) |\hat{\mathbf{p}}|^2 = \frac{2 \xi_2}{M_{\rm Pl}} |\hat{\mathbf{p}}|^2.
\end{equation}

Finally, for higher-order deformations ($n \ge 3$), the system enters an ultraviolet regime where the mass scaling inverts. The Hamiltonian becomes proportional to the inverse of the mass powers, scaling as $\hat{H} \propto (mc)^{2-n} |\mathbf{p}|^n$. For the cubic case ($n=3$), this yields:
\begin{equation}
    \hat{H}_{\text{LIV}}^{(3)} = \frac{\xi_3}{M_{\rm Pl} c} \left( \frac{1}{M_{\text{nuc}}} + \frac{1}{m_\ell} \right) |\hat{\mathbf{p}}|^3 \approx \frac{\xi_3}{M_{\rm Pl} m_{\ell} c} |\hat{\mathbf{p}}|^3.
\end{equation}
Since $M_{\text{nuc}} \gg m_\ell$, the nuclear contribution is heavily suppressed. Thus, for $n \ge 3$, the lepton becomes the primary probe of quantum gravity, and the sensitivity is driven by the smallness of the electron (or muon) mass, validating the single-particle approximation typically employed in high-energy analyses.

\subsection{First-Order Energy Shifts and the Master Equation}

We employ first-order time-independent perturbation theory to derive the theoretical expressions for the modified atomic spectrum, providing the necessary framework for a global statistical adjustment of the physical parameters. Instead of treating fundamental constants as fixed inputs, these spectral corrections enable simultaneous parameter estimation of the Rydberg constant shift ($\Delta R_y$), the proton charge radius ($r_{p}$), and the Lorentz-violating parameter ($\xi$) via a robust MCMC analysis. The energy shift for a state characterized by quantum numbers $\nu, l$ (principal and orbital) is given by
\begin{equation}
    \Delta E_{\nu, l}^{(n)} = \langle \nu l | \hat{H}_{\text{LIV}}^{(n)} | \nu l \rangle = \frac{\xi_n c^{2-n}}{M_{\rm Pl}} \sum_{i=a,b} \left( m_i^{2-n} \right) \langle |\hat{\mathbf{p}}|^n \rangle_{\nu l},
\end{equation}
where the momentum expectation value is computed through the momentum space integral
\begin{equation}
    \langle |\hat{\mathbf{p}}|^n \rangle_{\nu l} = \int d^3p \, p^n |\phi_{\nu l}(\mathbf{p})|^2,
\end{equation}
with $\phi_{\nu l}(\mathbf{p})$ representing the atomic wavefunction in momentum space. We highlight that parity symmetry forbids first-order shifts for vector momentum operators. Our analysis properly focuses on the magnitude of the momentum operator $|\hat{\mathbf{p}}|^n$, ensuring real, physical energy shifts consistent with quantum mechanical principles.

By combining the unified Hamiltonian derived in the center of mass frame with these expectation values, we derive the exact master equation for the LIV-induced energy shift. Unlike simplified single-particle models, this expression rigorously accounts for the reduced mass ($M_{\rm red}$) dynamics. Evaluating the integral analytically factors out the characteristic momentum scale of the state and yields
\begin{equation}
    \Delta E_{\nu, l}^{(n)} = \frac{\xi_n c^2}{M_{\rm Pl}} \left[ M_{\text{nuc}}^{2-n} + m_\ell^{2-n} \right] M_{\rm red}^n \left( \frac{Z \alpha}{\nu} \right)^n \mathcal{I}_{n, l},
    \label{eq:master_shift}
\end{equation}
where the dimensionless structural coefficient $\mathcal{I}_{n, l}$ completely characterizes the LIV deformation geometry for the specific state under consideration. The exact numerical values of this coefficient for the relevant states are directly summarized in Table \ref{tab:pmedioH}. To address the leading-order phenomenological effects, the analysis is restricted to the linear deformation case $n=1$, in which $\xi_1 \equiv \xi$. This approach expresses energy corrections in terms of the exact expectation values $\langle |\hat{\mathbf{p}}|\rangle_{\nu l}$, allowing a direct comparison between theoretical bounds and experimental data.



\section{MDR-Induced Energy Shifts in Atomic Spectra}\label{sec:dataset}

To carry out a comprehensive study of the influence of a modified
dispersion relation (MDR) induced by the IR/UV mixing of quantum-gravity effects on the hydrogen atom spectrum, we analyzed a set of hydrogen transitions measured with high precision in recent experiments. From the list of transitions used by CODATA22 \cite{Mohr:2024kco}, we selected those determined with the highest
precision, corresponding to uncertainties on the order of \textrm{kHz} (see Table \ref{table:1}).

In addition to the electronic hydrogen transitions, we included in our
analysis the measurement of the Lamb shift in muonic hydrogen, due to its high sensitivity to the proton charge radius. It is important to emphasize that the MDR effect on this transition is negligible, given the relatively low experimental precision compared to that of electronic hydrogen. Thus, the determination of the proton radius from this measurement is not affected by new physics within the accuracy currently achieved in muonic hydrogen spectroscopy.

The last item of Table \ref{table:1} refers to the determination of the Rydberg constant
from the analysis of transitions between Rydberg states of hydrogen. Although
this measurement presents a relatively large uncertainty, it is relevant to
our study because it is practically independent of both the proton radius and
MDR effects.

\begin{table}[h]
\centering%
\resizebox{\columnwidth}{!}{%
\begin{tabular}
[c]{l l l}\hline\hline
& \\[-1ex]%
Transition or energy interval & Experimental value (kHz) & Ref.\\[1.5ex]\hline
& \\[-1ex]%
$2S_{1/2}^{F=1}-1S_{1/2}^{F=1}$ & 2 466 061 102 474.806(10)$^{a}$ & \cite{Parthey:2011lfa}\\[1.5ex]
$2S_{1/2}^{F=1}-1S_{1/2}^{F=1}$ & 2 466 061 102 474.789(11)$^{a}$ & \cite{Matveev:2013orb}\\[1.5ex]
$3S_{1/2}^{F=1}-1S_{1/2}^{F=1}$ & 2 922 742 936 716.72(72)$^{b}$ & \cite{Grinin:2020txk}\\[1.5ex]
$3S_{1/2}^{F=1}-1S_{1/2}^{F=1}$ & 2 922 742 936 722.4(2.6)$^{c}$ & \cite{Fleurbaey:2018fih}\\[1.5ex]
$4P^{F=1}-2S_{1/2}^{F=0}$ & 616 521 064 178.9(2.3)$^{d}$ & \cite{Beyer:2017gug}\\[1.5ex]
$2S_{1/2}^{F=0}-2P_{1/2}^{F=1}$ & 909 871.7(3.2) & \cite{Bezginov:2019mdi}\\[1.5ex]
$8D_{5/2}-2S_{1/2}^{F=1}$ & 770 649 517 181.7(2.0)$^{e}$ & \cite{Brandt:2021yor}\\[1.5ex]
$E_{L}(\mu H)/h$ (Lamb shift in $\mu H$) & $48.932\,99(56)\times10^{9}$ & \cite{Antognini:2013txn}\\[1.5ex]
$cR_{\infty}$ (from Rydberg states) & 3 289 841 960 306(69) & \cite{devries2002precision, Pohl:2016glp}\\[1.5ex]
\hline\hline
\end{tabular}
}
\caption{Experimental values of transition frequency with their uncertanties
in kHz. \label{table:1}}
\vspace{5pt}
{\small \textit{Note:} The frequencies listed correspond to the specific hyperfine transitions. Where the cited references report unperturbed centroid frequencies, we have applied the respective hyperfine structure (HFS) shifts, denoted as $\Delta f_{\mathrm{HFS}}$, to reconstruct the physical transitions:
$^{(a)}$~$\Delta f_{\mathrm{HFS}} = -310\,712.229$ kHz. 
$^{(b)}$~$\Delta f_{\mathrm{HFS}} = -341\,949.070$ kHz. 
$^{(c)}$~$\Delta f_{\mathrm{HFS}} = -341\,949.077$ kHz. 
$^{(d)}$~Corresponds to the combination $\tfrac{2}{3}(4P_{3/2}^{F=1}-2S_{1/2}^{F=0}) + \tfrac{1}{3}(4P_{1/2}^{F=1}-2S_{1/2}^{F=0})$. The applied shift from the reported centroid is $\Delta f_{\mathrm{HFS}} = +132\,552.092$ kHz. 
$^{(e)}$~$\Delta f_{\mathrm{HFS}} = -44\,389.2$ kHz.}
\end{table}

To evaluate the effects of possible new physics, it is necessary to compare
the experimental values with the theoretical predictions of the model. In the
hydrogen atom, the stationary states are usually identified by four quantum
numbers $\left(  n,\ell,j,F\right)  $, where $n$ is the principal quantum
number, $\ell$ characterizes the orbital angular momentum, $j$ denotes the
total electronic angular momentum, and $F$ corresponds to the total atomic
angular momentum, including the nuclear spin. The energy of these states can
be expressed, including the contribution of the MDR term, as the sum of four
components:%
\begin{align}
E\left(  n,\ell,j,F\right)  &=-\frac{hcR_{\infty}}{n^{2}}\frac{M_{\rm red}}{m_{e}%
}+\frac{C_{\rm NS}}{n^{3}}r_{p}^{2}\delta_{\ell0} \label{E}\\
&+E_{\rm fs+hfs}\left(  n,\ell
,j,F\right)  +\xi E_{\rm MDR}(n,\ell).\nonumber
\end{align}

The first one is the dominant Bohr term, proportional to the Rydberg constant
$(R_{\infty})$; the second term, present only in states with zero orbital
angular momentum $(\ell=0)$, reflects the influence of the finite nuclear size
and depends on the proton charge radius ($r_{p}$); the third term contains the
combined contributions of the fine and hyperfine structure of the atom; and
the final component represents the correction introduced by new physics, whose
strength is regulated by the dimensionless parameter $\xi$. The constant $c$
denotes the speed of light, and $h$ is Planck's constant, $M_{e}$ is the
electron mass and $M_{\rm red}$ is the reduced mass of hydrogen.

The first three terms in (\ref{E}) are well known within the framework of Quantum Electrodynamics. On the other hand, the MDR contribution, treated as a small correction, can be estimated using the standard perturbative method. As we have previously discussed, in the first approximation order, $E_{\rm MDR}$ corresponds to the expectation value
of the MDR Hamiltonian, $\left\langle H_{\rm MDR}\right\rangle $, evaluated in the
unperturbed states of the dominant approximation. Thus, from Eq. \eqref{eq:master_shift}, we can write:%
\begin{align}
    E_{\rm MDR}\left(  n,\ell\right) &= \frac{\Delta E_{n, l}^{(1)}}{\xi} =\frac{M_{\rm proton}}{M_{\rm Pl}}c\left\langle
|\hat{\mathbf{p}}|\right\rangle _{n,\ell}\notag\\
&=\frac{M_{\rm proton}}{M_{\rm Pl}}cA_{n,\ell}\left(  \alpha
M_{\rm red}c\right).
\end{align}
In the Table \ref{tab:pmedioH}, we present the values of the coefficients $A_{n,\ell}$ for the states that will be analyzed in our
study.
\begin{table}[th]
\centering
\begin{tabular}[c]{lcccr}\hline\hline
& \\[-1ex]%
State &  $A_{n\ell}$ & $\langle|\hat{\mathbf{p}}|\rangle$(eV/$c$)  & $E_{\rm MDR}/h$(kHz)\\[1.5ex] \hline
& & & & \\[-1ex]%
$1S$ & $\dfrac{8}{3\pi}$ & 3163.50 &  $5.88\times10^{-5}$\\[1.5ex]%
$2S$ & $\dfrac{16}{15\pi}$ & 1265.40 & $2.35\times10^{-5}$\\[1.5ex]%
$2P$ & $\dfrac{64}{45\pi}$ & 1687.20 & $3.14\times10^{-5}$\\[1.5ex]%
$3S$ & $\dfrac{24}{35\pi}$ & 813.47 & $1.51\times10^{-5}$\\[1.5ex]%
$4P$ & $\dfrac{128}{225\pi}$ & 674.88 & $1.25\times10^{-5}$\\[1.5ex]%
$8D$ & $\dfrac{22165504}{80405325\pi}$ & 327.03 & $6.08\times10^{-6}$\\[1.5ex]%
\hline\hline
\end{tabular}
\caption{Average value of the magnitude of the linear momentum in the hydrogen atom and of the MDR energy correction $E_{\rm MDR}(n,\ell)$.}\label{tab:pmedioH}%
\end{table}

We aimed to examine the constraints on the parameter $\xi$ imposed by the available spectroscopic data. To achieve this, we perform a global adjustment of the parameters $R_{\infty}$, $r_{p}$, and $\xi$. The statistical inference is conducted using a likelihood function based on the chi-squared ($\chi^2$) statistic, which enables us to obtain robust marginalized confidence intervals through a Markov Chain Monte Carlo (MCMC) analysis.

It is important to emphasize that, in the adjustment procedure, the
experimental data were treated as statistically independent, except for the two measurements of the $1S-2S$ transition, which exhibit a correlation of $0.7069$ according to CODATA22 \cite{Mohr:2024kco}.

\begin{widetext}
\begin{table}[!htbp]
\hfill
\begin{minipage}{0.92\textwidth}
\small
\begin{tabular}[c]{lrrrrrr}\hline\hline\\[-1ex]
Transition & $a_{i}$ & $b_{i}$ & $c_{i}$ & $d_{i}$ & $e_{i}$ & $f_{i}$\\[1.5ex]\hline
&  &  &  &  &  & \\[-1ex]%
$2S-1S$ & $3/4$ & $-1369.025$ & $22679757.5$ & $-3.52716\times10^{-5}$ & $-7/8$ & 0\\[1.5ex]%
$3S-1S$ & $8/9$ & $-1506.948$ & $30731931.3$ & $-4.36696\times10^{-5}$ & $-26/27$ & 0\\[1.5ex]%
$4P-2S$ & $3/16$ & $-195.575$ & $11458397.7$ & $-1.09734\times10^{-5}$ & $-1/8$ & 0\\[1.5ex]%
$2S-2P_{1/2}$ & 0 & $195.575$ & $909735.7$ & $-7.83814\times10^{-6}$ & $1/8$ & 0\\[1.5ex]%
$8D-2S$ & $15/64$ & $-195.575$ & $12509911.4$ & $-1.74373\times10^{-5}$ & $-1/8$ & 0\\[1.5ex]%
Lamb-shift $\left(  \mu H\right)  $ & 0 & $-1.263617\times10^{9}$ & $49.82588\times10^{9}$ & $+1.457418\times10^{-3}$ & 0 & 1\\[1.5ex]%
prior: $\Delta_{1S}=0$ & 0 & 0 & 0 & 0 & 1 & 0\\[1.5ex]%
prior: $\Delta_{L\mu H}=0$ & 0 & 0 & 0 & 0 & 0 & 1\\[1.5ex]%
$R_{\infty}\left(  \text{Ryd-states}\right)  $ &  $M_{e}/M_{\rm red}$ & 0 & 0 & 0 & 0 & 0\\[1.5ex]\hline\hline
\end{tabular}
\caption{Theoretical expression for transitions frequencies in H: $\nu_{i}^{\mathrm{th}}=a_{i}(M_{red}/M_{e})cR_{\infty}+b_{i}r_{p}^{2}+c_{i}+d_{i}\xi+e_{i}\Delta_{1S}+f_{i}\Delta_{L\mu H}$ utilized in the least-squared fit.\label{table:3}}
\end{minipage}
\hfill
\end{table}
\end{widetext}

Another relevant aspect is that the ratio $M_{\rm red}/M_{e}$ was included in our analysis as an input parameter. The value employed in our calculations was obtained from the proton-to-electron mass ratio:
\[
M_{\rm proton}/M_{\rm electron}=1836.152673346(18),
\]
as determined in Penning trap experiments \cite{2017PhRvL.119c3001H}. The
uncertainty of this measurement propagates into an error smaller than $0.2$ kHz in the determination of the Rydberg constant, which is therefore negligible compared to the final uncertainty obtained in the adjustments.

With regard to the contributions of the fine and hyperfine structure of the atom, it should be emphasized that, strictly speaking, these terms also depend on the Rydberg constant. However, given the current experimental limitations, it is not necessary to calculate $E_{\rm fs+hfs}$ using the most precise value of $R_{\infty}$. For this reason, the fine and hyperfine contributions can be determined separately, employing the values of the fine-structure constant, the electron mass, and other relevant quantities measured in independent experiments. In this work, we used the data from \cite{Pohl:2016glp}, for the fine-structure terms. The hyperfine contribution, in turn, was taken from Ref. \cite{2016PhRvA..93b2513H}.

Analogously, the coefficient $C_{\rm NS}$, associated with the energy correction due to the finite nuclear size, was determined using values of the fundamental constants extracted from independent sources. For hydrogen, the value of this coefficient, at a precision compatible with the experimental uncertainties, is given by $C_{\rm NS}/h=1564.60$ kHz/fm$^{2}$ \cite{Pohl:2016glp}, where we present the factor divided by Planck's constant, so that the contribution is expressed in units
of kHz.

\begin{widetext}
\begin{table}[!htbp]
\hfill
\begin{minipage}{0.92\textwidth}
\small
\begin{tabular}{lccccc}
\hline\hline \\[-1ex]
Model/Transitions & $cR_{\infty} - 3289841960000$ (kHz) & $r_{p}$ (fm) & $\xi\;(\times10^{5})$ & $\chi_{\mathrm{red}}^{2}$ & $\nu$ \\[1.5ex]\hline\\[-1ex]
1) MDR/Complete & $218.3(11)$ & $0.84065(39)$ & $-6.5(2.1)$ & $3.566
$ & 6\\[1.5ex]
2) MDR/ without $E_{L}[\mu H]$ & $294.7(42)$ & $0.85295(641)$ & $1.8(4.8)$ & $3.549
$ & 5\\[1.5ex]
3) MDR/ without $\mu H$ and Ryd-States & $288.2(52)$ & $0.85200(793)$ & $1.1(6.0)$ & $4.425
$ & 4\\[1.5ex]
4) MDR/ $1S-2S + E_{L}[\mu H]\, +$ Ryd-States & $306.0(69)$ & $0.84060(39)$ & $12.3(14.7)
$ & $4.413
$ & 1\\[1.5ex]
5) MDR/ only Lamb Shift ($H$ and $\mu H$) & $-$ & $0.84060(39)$ & $2.8(4.1)$ & $-$ & 0\\[1.5ex]
6) no MDR/Complete & $251.9(3)$ & $0.84079(38)$ & 0 & $4.390
$ & 7\\[1.5ex]
7) no MDR/without $\mu H$ and Ryd-States & $279.0(8)$ & $0.85067(282)$ & 0 & $3.547
$ & 5\\[1.5ex]\hline\hline
\end{tabular}
\caption{Parameter adjustments from Hydrogen spectroscopy for different experimental configurations where $cR_{\infty}=3289841960(\mathrm{xxx})$ kHz and the first column shows the last digits relative to the base value.}
\label{table:adjustments}
\end{minipage}
\hfill
\end{table}
\end{widetext}

Another aspect to be considered in the adjustment procedure is that the
corrections arising from the fine structure carry theoretical uncertainties.
The largest of these uncertainties occurs in the $1S$ state and amounts to $2.5$ kHz \cite{Pohl:2016glp}. For the other $nS$ states, the theoretical uncertainties are strongly correlated with the $1S$ uncertainty and scale approximately as $2.5/n^{3}$. In contrast, the theoretical uncertainties of states with nonzero orbital angular momentum, as well as those of the hyperfine structure, are relatively small and can be neglected in our discussion.

Following the procedure adopted by CODATA22 \cite{Mohr:2024kco}, we treat the theoretical uncertainties as additional parameters to be adjusted together with $R_{\infty},r_{p}$ and $\xi$. Thus, we introduce two new parameters, $\Delta_{1S}$ and $\Delta_{L\mu H}$, which represent, respectively, the uncertainty of the $1S$ state in hydrogen and the theoretical Lamb-shift uncertainty in muonic hydrogen. These extra parameters must be properly incorporated into the theoretical expressions, leading to the following modifications:%
\begin{align}
E^{\prime}\left(  n,\ell,j,F\right)   &  =E\left(  n,\ell,j,F\right)
+h\frac{\Delta_{1S}}{n^{3}}\delta_{\ell0}.\\
E_{L}^{\prime}\left(  \mu H\right)   &  =E_{L}\left(  \mu H\right)
+h\Delta_{L\mu H}.
\end{align}
To include these extra parameters in the adjustment process, we introduce two new equations given by $\Delta_{1S}=0.0(2.5)$ kHz and $\Delta_{L\mu H}=0.0(6.0)\times
10^{5}$ kHz. These equations indicate our prior expectation that both
parameters vanish, while accounting for their respective uncertainties. In the Table \ref{table:3}, we explicitly describe the theoretical expressions employed in the fitting procedure.

\section{Constraints from hydrogen spectroscopy}
\label{sec:bounds}
The adjustment of the parameters was carried out by comparing the theoretical expressions in Table~\ref{table:3} with the experimental values of the transition frequencies presented in Table~\ref{table:1}. In our study we analyzed several configurations each composed of different sets of transitions as summarized in Table~\ref{table:adjustments}.

\subsection{Statistical Methodology and MCMC Analysis}
\label{subsec:mcmc_methodology}

In order to evaluate the statistical significance of the MDR effects and to map the degeneracies among the physical parameters, we employ a Markov Chain Monte Carlo (MCMC) algorithm \cite{press2007numerical}. Since the model is linear in the fitted parameters, the minimum and the covariance matrix are obtained in closed form, and the MCMC serves to map the parameter degeneracies and to marginalize over the nuisance parameters. We construct the likelihood function based on a standard Gaussian distribution where the total $\chi^2$ accounts for the experimental transition frequencies and their corresponding uncertainties. Given the persistent underestimation of experimental errors in global hydrogen adjustments, we adopt the procedure of globally rescaling the uncertainties by the Birge factor $R_B = \sqrt{\chi^2_{\mathrm{red}}}$.

In our MCMC runs we assume flat priors for the physical parameters $\Delta R_y$, $r_p$, and $\xi$. The theoretical uncertainties are treated as nuisance parameters with Gaussian priors centered at zero, specifically $\Delta_{1S} = (0.0 \pm 2.5)$ kHz and $\Delta_{L\mu H} = (0.0 \pm 6.0) \times 10^5$ kHz. The chains are run until full convergence is achieved to ensure the posterior distributions are reliably sampled.

\subsection{Global Fit and Parameter Constraints}
\label{subsec:global_fit}

To test the viability of the Lorentz-violating model, we quantify the improvement of the fit by comparing the minimized $\chi^2$ of the MDR parametrization against the null hypothesis of standard quantum electrodynamics. According to the results for the complete dataset shown in Table~\ref{table:adjustments}, the null hypothesis of zero Lorentz violation (Adjustment 6) yields a reduced chi-square of $\chi^2_{\mathrm{red}} = 4.390$ for 7 degrees of freedom, while the inclusion of the LIV parameter $\xi$ (Adjustment 1) lowers it to $\chi^2_{\mathrm{red}} = 3.566$ for 6 degrees of freedom, an absolute improvement of $\Delta \chi^2 \approx 9.34$ over the standard physics scenario. In this first configuration, which includes all transitions from Table~\ref{table:1}, the adjustment would exclude $\xi = 0$ at the $3\sigma$ level. However, due to the high value of the reduced chi-squared obtained in this adjustment, such a conclusion is not statistically significant. The data set is mutually inconsistent whether or not the Lorentz-violating term is included, with goodness-of-fit probabilities of $7.0\times10^{-5}$ and $1.6\times10^{-3}$ respectively, so that the inclusion of $\xi$ does not make the adjustment acceptable and the apparent improvement cannot be read as evidence for new physics. Once the uncertainties are rescaled by the Birge factor, as described below, the same improvement reads $\Delta\chi^2/R_B^2 = 2.62$ and the significance drops to $1.6\sigma$. This interpretation is reinforced by the analysis of the other configurations, which yield discrepant results for the best-fit value of the parameter $\xi$. In configurations ``2'' and ``3'', for example, where the muonic hydrogen measurement was excluded, the optimal value of the parameter is positive and consistent with $\xi = 0$. This indicates that the preference for a negative $\xi$ in the complete adjustment originates from the tension between the muonic and electronic determinations of $r_p$, so that $\xi = 0$ remains statistically viable.

It is important to emphasize that the persistently high chi-squared values are not exclusive to the MDR framework. As noted by CODATA22 \cite{Mohr:2024kco}, standard adjustments without statistical corrections yield $\chi^2$ values with a mere 0.001\% probability of occurring due to pure statistical fluctuations. For instance, imposing $\xi=0$ on restricted datasets (Adjustment 7) reproduces the exact CODATA22 \cite{Mohr:2024kco} parameters recommendations for purely electronic data, yet the $\chi^2_{\mathrm{red}}$ remains high. Furthermore, in configurations where the muonic hydrogen measurement is excluded (Adjustments 2 and 3), the optimal value for the LIV parameter turns positive and is consistent with $\xi=0$ within $1\sigma$, though large $\chi^2_{\mathrm{red}}$ values persist. This overarching behavior strongly justifies our adoption of the Birge factor to globally rescale uncertainties prior to the MCMC analysis.

We also performed additional adjustments in which we removed each measurement in Table~\ref{table:1} in turn. Removing the muonic hydrogen input changes the best fit to $\xi = +1.8(4.8)\times10^{5}$ and simultaneously moves the proton radius from $0.84065(39)$ fm to $0.85295(641)$ fm, whereas removing any other single measurement leaves $\xi$ negative, in the range $-3.4$ to $-9.6$ in units of $10^{5}$. With $r_p$ anchored by the muonic measurement, the electronic data, which on their own favor a larger radius, require an $n$-dependent shift in the $S$ states, and $\xi$ supplies it. The parameter therefore absorbs part of the known proton-radius tension.

The MCMC analysis confirms the best-fit values obtained via gradient-based minimization to the complete Adjustment 1. The algorithm yields $\Delta R_y = 218 \pm 11$ kHz and $r_p = 0.84065 \pm 0.00039$ fm alongside a Lorentz-violating parameter of $\xi = (-6.5 \pm 2.1)\times10^{5}$ at the 68.3\% confidence level. The Birge factor is evaluated at the minimum of the complete fit, $R_B = \sqrt{\chi^2_{\mathrm{red}}} = 1.888$, and applied to the full covariance matrix of the input data, including the two theoretical nuisance priors and the correlation between the two $1S$--$2S$ measurements. Because the model is linear in the fitted parameters, this leaves all best-fit values unchanged and enlarges every parameter uncertainty by the single factor $R_B$, so that the uncertainty of $\xi$ increases from $2.12\times10^{5}$ to $4.01\times10^{5}$. The resulting 95\% confidence level interval for the MDR parameter is
\begin{equation}
\xi\in\left[  -14.3,\;1.37\right]  \times10^{5},
\end{equation}
which coincides with the profile-likelihood interval. This is the most robust constraint that can be extracted from the most recent hydrogen spectroscopic data. With the rescaled uncertainties, $\xi = 0$ lies $1.6\sigma$ from the best fit, so that the standard physics scenario is comfortably contained within the $2\sigma$ region.

\subsection{The Role of Muonic Hydrogen in Breaking Degeneracies}
\label{subsec:degeneracies}

\begin{figure}[htbp]
    \centering
    \includegraphics[width=1\linewidth]{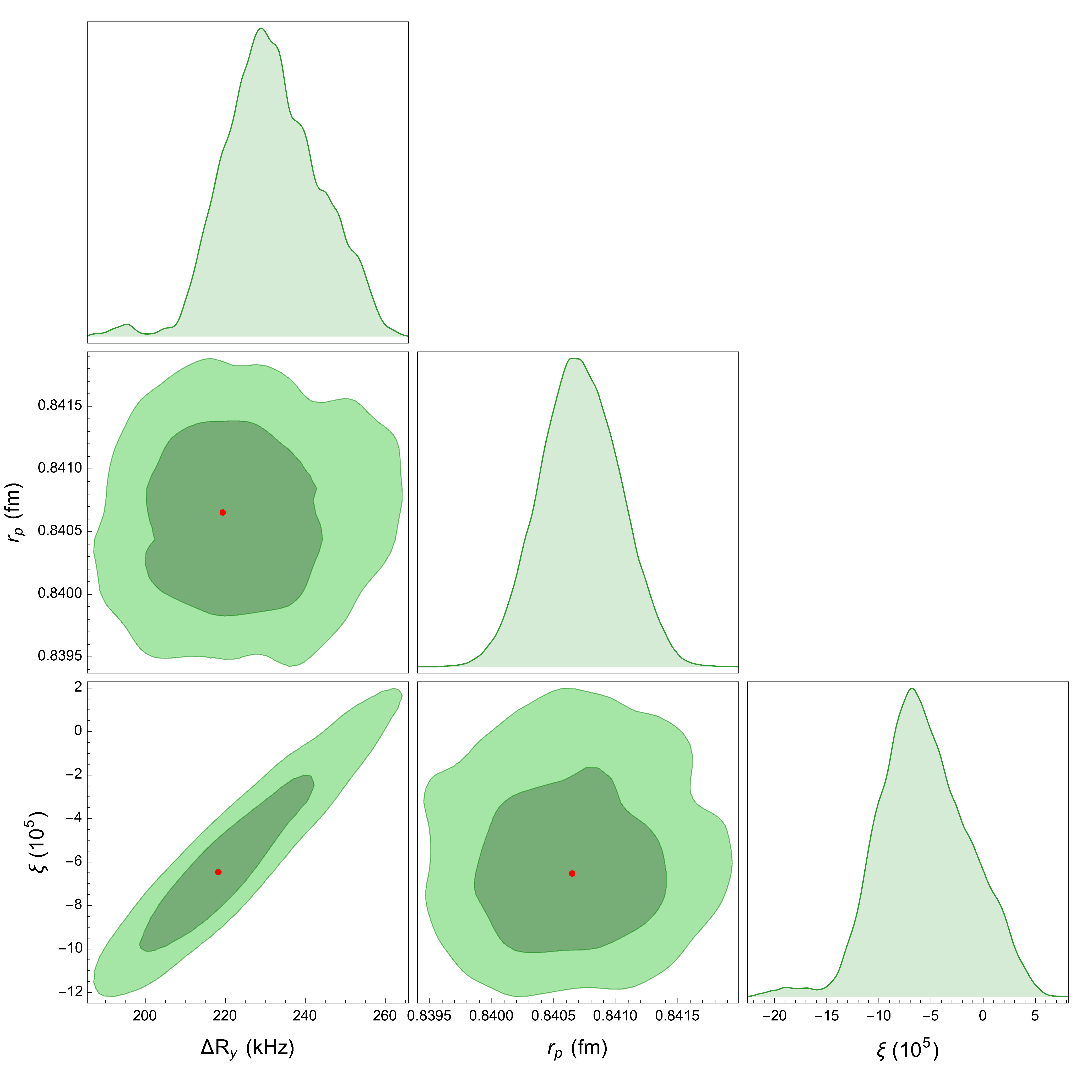} 
    \caption{One-dimensional marginalized posterior distributions and two-dimensional joint confidence contours for the Rydberg constant shift ($\Delta R_y$), the proton radius ($r_p$), and the LIV parameter ($\xi$). The contours are shown at the $1\sigma$ and $2\sigma$ levels and the red dots indicate the best-fit values.
    }
    \label{fig:corner}
\end{figure}

To better visualize the degeneracies among the fitted parameters, we present the corner plot in Fig.~\ref{fig:corner}. This figure displays the one-dimensional marginalized posterior distributions on the diagonal and the two-dimensional joint confidence contours at the $1\sigma$ and $2\sigma$ levels on the off-diagonal panels.
The two-dimensional contours reveal critical physical insights regarding the structure of the theory. A strong positive correlation is observed between $\Delta R_y$ and $\xi$ in the bottom-left panel. This degeneracy indicates that the theoretical framework attempts to accommodate the LIV modification by artificially shifting the value of the Rydberg constant. Because $\xi$ and $\Delta R_y$ can partially compensate for each other in the energy expressions for the $S$-states, this strong correlation emphasizes the necessity of the MCMC approach to properly marginalize the parameters and estimate the true uncertainty of $\xi$.

In contrast, the proton radius $r_p$ exhibits almost no correlation with $\xi$ and only a weak correlation with $\Delta R_y$. This behavior departs from the well-known degeneracy between the Rydberg constant and the proton radius found in purely electronic hydrogen spectroscopy \cite{Beyer:2017gug, Mohr:2024kco}. This decoupling occurs because $r_p$ is robustly anchored by the muonic hydrogen Lamb shift measurement, which is virtually immune to the MDR effects at the current level of experimental precision. This structural independence indicates that the constraint on $\xi$ is completely isolated from the proton radius uncertainty, and it gives exceptional robustness to our final bound.

\subsection{Testing Rydberg and Lamb Shift Configurations}
\label{subsec:data_subsets}
Complementing the analysis, additional configurations were investigated.
Adjustment 4, for instance, can be interpreted as a direct test of
modifications to the dispersion relation through hydrogen spectroscopy, using the $1S-2S$ transition, which is experimentally the most precise. This interpretation follows from the fact that, in this adjustment, we avoid the influence of MDR on the determination of both the Rydberg constant and the proton radius. As previously mentioned, the measurement of $R_{\infty}$ from transitions between Rydberg states of the atom is essentially free from proton-radius effects. Moreover, since the principal quantum number is high, the MDR contribution to these states is negligible compared to its impact on the $1S$ and $2S$ levels.

Another relevant aspect of this configuration is that the proton radius is determined almost exclusively from the muonic hydrogen Lamb-shift measurement, in which MDR has practically no measurable effect. Indeed, although MDR effects are theoretically stronger in muonic hydrogen than in electronic hydrogen, any potential signal of the modified dispersion relation would be obscured by the experimental uncertainty of the Lamb-shift measurement in $\mu H$, which is of the order of $5.6$ \textrm{GHz}.

The main limitation of Adjustment 4, however, lies in the relatively poor
precision of the determination of $R_{\infty}$ by transitions between Rydberg states. Nevertheless, it is worth emphasizing that there are promising prospects for improvements in this measurement \cite{Jentschura:2008zz, Hansch:2006bi}, which could make
this test significantly more sensitive in the future.

Configuration 5 is another adjustment that we highlight in our study, as it is based on the analysis of the same $2S-2P$ transition in both electronic and muonic atoms. Another advantage of this subset is that it also avoids the influence of the Rydberg constant measurement, since the levels share the same principal quantum number. The results from the adjustment of this configuration, when analyzed separately, provides comparatively the strongest constraints on the MDR parameter.

\subsection{Macroscopic Scaling and Future Prospects}
\label{subsec:macroscopic_scaling}

It is instructive to compare our results with tests of MDR effects performed in other experiments. In Ref. \cite{Amelino-Camelia:2025zkq}, for instance, it was suggested that the tensions between different determinations of the fine-structure constant can be explained by quantum-gravity--induced MDR effects. The discrepancy arises between the value of $\alpha$ obtained from the electron's gyromagnetic factor combined with rubidium interferometry, and the value derived from cesium atom interferometry. MDR effects have negligible influence on the gyromagnetic factor and naturally cancel out in the specific experimental setup of rubidium interferometry, whereas the cancellation does not occur in the cesium experiment. Resolving this discrepancy through Lorentz violation requires an MDR parameter of approximately $\xi[^{133}\text{Cs}] \approx 0.6$ \cite{Amelino-Camelia:2025zkq}. Our notation was chosen in order to emphasize that the inferred value for the
MDR parameter was obtained from experiments with cesium atoms, whereas our result is based on hydrogen.

In principle, these values may differ due to the so-called soccer ball problem, which concerns the need to suppress MDR effects in macroscopic systems. One way to circumvent this issue is to assume that the MDR-parameter scales linearly with the number of the elementary constituents of the composite body \cite{Hossenfelder:2014ifa}. Under this scaling hypothesis, the value inferred from the $^{133}\text{Cs}$ atom would correspond to an effective parameter of approximately $133 \times 0.6 \approx 80$ for the hydrogen atom.

Thus, the modification of the dispersion relation would lead to a frequency shift of approximately $3$ Hz in the hydrogen $1S-2S$ transition. This result shows that the current experimental precision of $10$ Hz  in measuring this transition is already very close to the sensitivity required to probe Planck-scale physics through MDR effects. However, due to the limitations in the independent determinations of the Rydberg constant, the proton radius and theoretical uncertainties, such LIV effects would still remain hidden, preventing the hydrogen spectroscopy from being fully conclusive as a test of the mixing IR/UV quantum-gravity effects.

\section{Conclusion}\label{sec:conc}

We have considered the effect of Lorentz invariance violation on the energy spectrum of electronic and muonic hydrogen atoms. The LIV effects were modeled by a modification of the dispersion relation, suppressed by a scale related to quantum gravity identified with the Planck mass. We have carried out a detailed analysis of several transitions, considering the important role of the theoretical and experimental uncertainties concerning the determination of the Rydberg constant and the proton radius. 

Our statistical inference allowed us to set a constraint on the dimensionless Lorentz violating parameter of the order $\sim 10^{5}$ at $95$\% CL, if the effect is suppressed by the Planck scale. The most important property of this bound is the fact that it surrounds the Planck scale and was derived using the simplest atoms that could be considered, which are the electronic and muonic hydrogen atoms. This alleviates possible ambiguities that could be present when analyzing more complex quantum systems, such as heavier atoms. The role of the uncertainties in the Rydberg constant and in the proton radius was also discussed. This analysis demonstrates that a more precise determination of the Rydberg constant and the proton radius  would allow atomic spectroscopy to fully scrutinize the Planck scale.

\bibliographystyle{utphys}
\bibliography{refs_lamb}

\end{document}